\documentclass[english]{article}

\usepackage{graphicx}
\usepackage{hyperref}
\usepackage{authblk}
\usepackage{float}
\usepackage{geometry}
\usepackage[backend=biber, sorting=none]{biblatex}
\usepackage{multicol}
\usepackage{amsmath}
\usepackage{bm}% bold math
\usepackage{braket}
\usepackage[utf8]{inputenc}

\usepackage{amsmath}
\usepackage{float}
\usepackage{graphicx}% Include figure files
\usepackage{bm}% bold math
\usepackage{braket}

\begin{document}

\title{All-Fiber Source and Sorter for Multimode Correlated Photons}

\author[1]{Kfir Sulimany}
\author[1 *]{Yaron Bromberg}

\affil[1]{Racah Institute of Physics, The Hebrew University of Jerusalem, Jerusalem 91904, Israel}
\affil[*]{Corresponding authors: yaron.bromberg@mail.huji.ac.il}

\date{}

\maketitle

\begin{abstract}
Photons occupying multiple spatial modes hold a great promise for implementing high-dimensional quantum communication. We use spontaneous four-wave mixing to generate multimode photon pairs in a few mode fiber. We show the photons are correlated in the fiber mode basis using an all-fiber mode sorter. Our demonstration offers an essential building block for realizing high-dimensional quantum protocols based on standard, commercially available fibers, in an all-fiber configuration.
\newline
\end{abstract}
\begin{multicols}{2}
\section{Introduction}
High-dimensional quantum bits hold great potential for quantum communication owing to their robustness to a realistic noisy environment \cite{cerf2002security, erhard2020advances, cozzolino2019high}. 
Implementations based on encoding information in the transverse spatial modes of photons are especially promising due to the large Hilbert space they span \cite{molina2007twisted, erhard2018twisted}.
In recent years, such implementations were successfully demonstrated in free-space \cite{krenn2015twisted, sit2017high}. Meanwhile, efforts for multimode fiber-based technologies are expected to achieve high-dimensional quantum communication without line of sight, based on existing multimode fiber components and infrastructures
 \cite{loffler2011fiber,kang2012measurement,cozzolino2019air,alarcon2021few,valencia2020unscrambling,cao2020distribution,zhou2021high}. 

The leading approach for generating entangled photons in transverse spatial modes is through spontaneous parametric down conversion in bulk crystals \cite{mair2001entanglement}.  However, it is extremely challenging to couple transverse entangled photons to fibers since it requires a precise mapping between the free space transverse modes and the fiber guided modes. Indeed, most demonstrations of distributing spatially entangled photons with fibers are limited to coupling of only two guided modes \cite{loffler2011fiber,kang2012measurement,cozzolino2019air,alarcon2021few}. Recently, distribution of a photon which is entangled in six spatial modes over a 2 meter-long fiber \cite{valencia2020unscrambling}, and in three spatial modes over a 1 km-long fiber were demonstrated \cite{cao2020distribution}. However, these methods require accurate calibrations, limiting implementations in real-life scenarios.

An alternative for coupling free-space entangled photons to a fiber, is to generate the photons inside the fiber by using spontaneous four-wave mixing (SFWM). Over the past two decades, generation of photon pairs by SFWM was studied using multiple types of single mode optical fibers \cite{park2020telecom}, including photonic crystal fibers \cite{sharping2004quantum, rarity2005photonic, fan2005efficient, cohen2009tailored}, dispersion shifted fibers \cite{li2005optical, takesue20051, dyer2009high}, and birefringent fibers \cite{smith2009photon, lugani2020spectrally}.
SFWM in multimode fibers was recently utilized for generating photons occupying a high dimensional transverse mode  \cite{cruz2016fiber, rottwitt2018photon, guo2019generation}. Generating photon pairs in a superposition of multiple fiber modes requires precise  analysis of the phase matching conditions that will allow multiple SFWM processes in the same spectral channel \cite{pourbeyram2018photon, rottwitt2019quantum, ekici2020graded, goudreau2020theory}. These theoretical works predict that the photon pair sources can be tunable for a wide range wavelengths, from the ultraviolet to the infra red and the telecommunication range. Experimentally, such phase matching conditions were recently studied for parametric amplification of weak signals \cite{shamsshooli2020toward}, but not in the spontaneous regime. Hence correlations between pairs of photons generated in multiple fiber modes were not measured to date.

In this work, we propose and demonstrate a fiber source of photon pairs which occupy multiple fiber modes. Our measurements prove that the photons are correlated in the guided mode basis, by mapping the modes the photons occupy to their arrival times at the end of a 1 km-long fiber. The 1 km-long fiber acts as an all-fiber in-line mode sorter, in contrast to bulk free-space mode sorters that are typically used for measuring correlations between transverse modes \cite{fontaine2019laguerre, mair2001entanglement, krenn2014generation, berkhout2010efficient}. 
Our in-line mode sorting configuration allows us to measure the two-dimensional (2D) histogram of the arrival times of the photons, which reveals that the photons occupy three guided modes of the fiber. By analyzing the histogram we achieve the two-photon modal decomposition and verify the spatial correlations of photon pairs generated in the multimode fiber.

\section{Results}
\subsection{Multimode Correlated Photons Source}
\begin{figure}[H]
\begin{centering}
\includegraphics[width=\columnwidth]{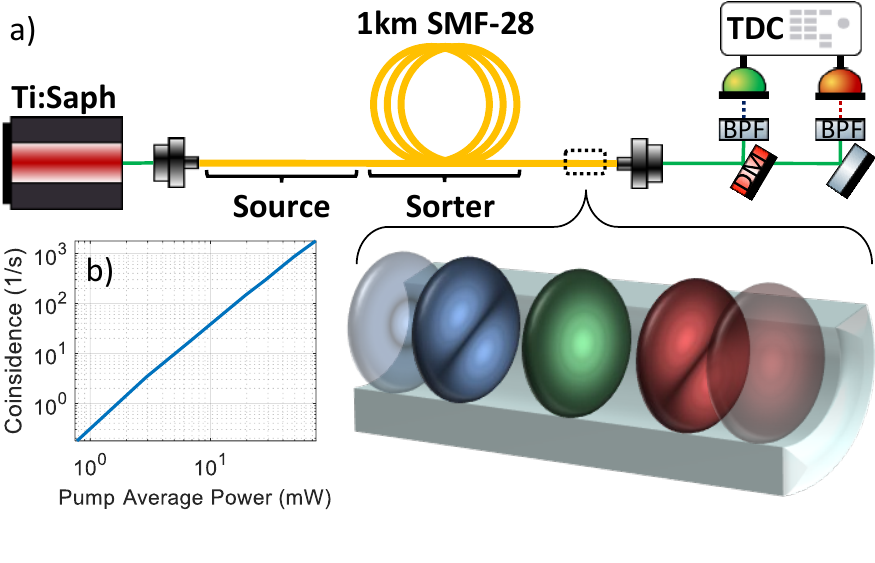}
\par\end{centering}
     \caption{ \textbf{An all-fiber multimode source and mode sorter for photon pairs correlated in the fiber modes.} a) Ultrashort pulses of 140fs $(\lambda_{pump}=695nm$) are coupled into a 1 km-long fiber. Pump photons are spontaneously annihilated and pairs of signal and idler photons are generated at two different spectral channels $(\lambda_{s}=542nm,\lambda_{i}=970nm)$. At these wavelengths, the fiber (SMF-28) supports a few modes, where the modal distribution of the photon pairs is determined by the phase matching condition of the fiber. After the first few tens of centimeters, the temporal spread of the pump pulse prevents SFWM. In the next 1-km long section of the fiber, the different modes are separated due to modal dispersion (inset). Higher spatial modes arrive after lower spatial modes, and shorter wavelengths arrive after longer wavelengths. At the output of the fiber the signal and idler photons are spectrally separated by a dichroic mirror (DM), filtered by a bandpass filter (BPF) and their arrival times are registered using two single photon detectors and a time-to-digital converter (TDC). An electronic delay of $70 ns$ is introduced to the idler detector to compensate for chromatic delay between the signal and idler photons. b) The experimentally measured coincidence rate as a function of the pump average power for a 20nm long fiber exhibiting quadratic scaling.}
 \label{Fig1}
 \end{figure}

Our source is based on coupling Ti:Sapphire mode-locked pulses (pulse duration $140 fs$, wavelength $\lambda_{pump}=695 nm$) into a few mode fiber as shown in Figure \ref{Fig1}. In SFWM, two pump photons are spontaneously annihilated, and two photons called signal and idler are generated in two spectral channels ($\lambda_{s}=542 nm$, $\lambda_{i}=970 nm$). Each spectral channel is composed of many different spatial modes. The photons occupy the guided modes of the fiber, which can be approximated by the linearly polarized (LP) modes of a weakly guiding optical fiber. The state of the photons is  determined by the phase matching conditions and can be written as: $|\Psi\rangle = \alpha|LP_{02}\rangle_{s}|LP_{01}\rangle_{i}+\beta|LP_{11}\rangle_{s}|LP_{11}\rangle_{i}$ where subscripts s (i) mark the mode of the signal (idler) photon and the coefficients $\alpha, \beta$ are determined by the nonlinear overlap integral (see Supplementary Equation 3). The term $|LP_{01}\rangle_{s}|LP_{02}\rangle_{i}$ is not present in the quantum state as the mode $LP_{02}$ is not guided in our fiber for the wavelength of the idler photon. Extension of this scheme to higher dimensions and other spectral bands is presented in the Supplementary Note 1.
 
The photon pairs are generated mostly in the first few tens of centimeters of the fiber, after which the peak power of the pump pulse is too weak for SFWM due to its temporal spreading, for more information see Supplementary Note 2. To quantify the efficiency of the pair generation we use a $20 cm$ section of SMF-28 to measure the coincidence detection rate as a function of the average pump power, exhibiting a quadratic scaling as expected for a four-wave mixing process (Fig. 1b). The coincidence to accidental ratio we obtain for a pump power of 10mW is 850 (see Supplementary Note 3 for more details). In principle, to improve the coincidence rate we could use higher pump powers. Increasing the pump power, however, will also increase parasitic Raman scattering. In our system Raman scattering hardly adds noise since it is temporally separated from the generated photon pairs. However, the pump power is limited since the photon counts due to Raman scattering exceed the maximal count rate of our detectors ($\approx5Mhz$). This limitation can be circumvented by using superconducting nanowire detectors with an order of magnitude higher maximal count rates ($\approx50Mhz$), or by using in-line fiber Bragg gratings to filter the pump light before the sorter, so that the pump will not generate Raman scattering along the 1 km-long fiber. 

\subsection{Multimode Photons Sorter}
Next, we use a 1 km-long section of the same fiber, which serves as a photon pairs source and as a mode sorter of the fiber’s guided modes. Due to modal Group Delay Dispersion (GDD), the arrival times of the photons at the end of the fiber depend on their modal distribution and their spectral channel, as depicted in Figure \ref{Fig1}. We can therefore map the arrival times of the photons to their modal decomposition, up to modal degeneracy in symmetric fiber cores. Although this sorting scheme is quite common in classical optics \cite{painchaud1992time}, it was only recently demonstrated at the single photon level for weak coherent pulses \cite{chandrasekharan2020observing}. Here we use the same principle for entangled photons. In our set-up, the temporal resolution is limited by the jitter of the avalanche photo diodes which is $400 ps$. Since the GDD of our fiber is in the scale of $1 ns/km$, a 1 km-long fiber is sufficient to temporally separate the modes. 

\subsection{Two Photon Modal Distribution Measurement}
\begin{figure}[H]
\begin{centering}
\includegraphics[width=\columnwidth]{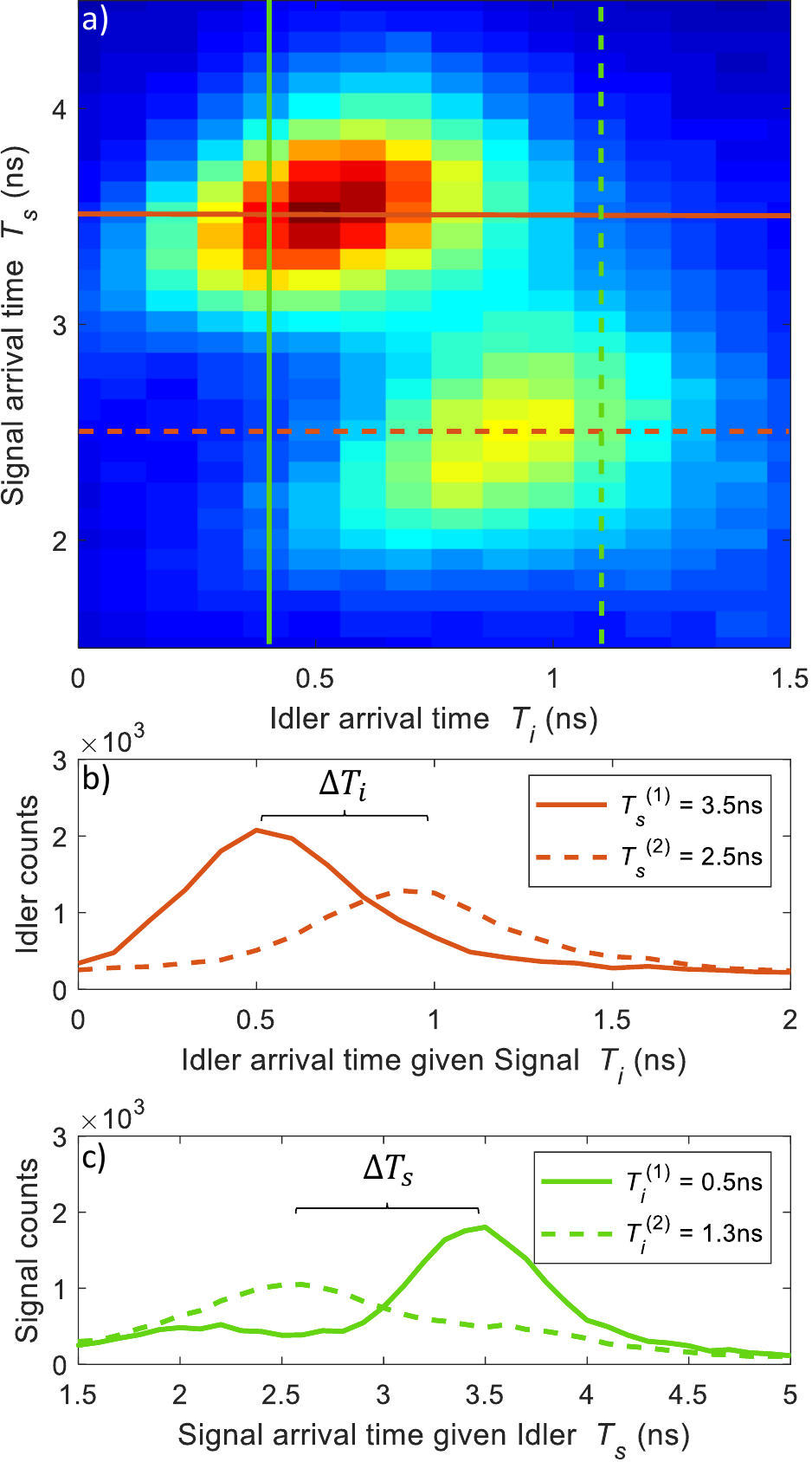}
\par\end{centering}
    \caption{\textbf{Temporal two-photon probability.} (a) Histogram of the arrival times $T_{s},T_{i}$ of the signal and idler photons. The arrival times are measured relative to an electronic trigger from the pump laser which serves as a global clock, and after adding an electronic delay of 70ns to the idler detector to compensate for the chromatic delay between the signal and idler photons. The two off-diagonal peaks indicate that the two-photon state is not separable. We therefore conclude that the photons are correlated in the modal basis. The two peaks correspond to occupation of modes $|LP_{02}\rangle_{s}|LP_{01}\rangle_{i}$ and $|LP_{11}\rangle_{s}|LP_{11}\rangle_{i}$, as verified by numerical computation of the fiber’s modal group delays. (b),(c) Cross sections of the two-dimensional histogram along the lines marked in (a), emphasizing the modal correlations. For example, post selecting events with an idler’s arrival time of $T_{i}^{(1)}=0.4ns$ (green solid curve) shows localization of the signal photon at $T_{s}^{(2)}=3.5ns$. In (b) the post selection is on the signal photon, while in (c) it is on the idler photon. The measured delay between the two peaks is $\Delta T_{s}=1ns$ for the signal photons and $\Delta T_{i}=0.5ns$ for the idler photons.}
\label{Fig2}
\end{figure}

To investigate the modal distribution of the two-photon state, we use the mode-to-time mapping and study the temporal two-photon probability $P(T_{s},T_{i})$ that describes the probability to detect a signal photon at time $T_{s}$ and an idler photon at time $T_{i}$. To this end, we plot the two-dimensional histogram of the arrival times after compensating for chromatic dispersion (Figure \ref{Fig2}(a)). Two correlation peaks are observed, corresponding to the delay between either $|LP_{02}\rangle_{s}$ and $|LP_{01}\rangle_{i}$ or between $|LP_{11}\rangle_{s}$ and $|LP_{11}\rangle_{i}$. Clearly, the two-photon probability is not-separable, indicating that photons are correlated in the modal basis.
To quantify the correlation of the two photons we post select two arrival times for the signal $(T_{s}^{(1)},T_{s}^{(2)})$ and two arrival times for the idler $(T_{i}^{(1)},T_{i}^{(2)})$. The post-selected arrival times are chosen to maximize the Pearson correlation coefficient: $PCC=\sum_{k=1}^{2} \sum_{l=1}^{2} P(T_{s}^{(k)},T_{i}^{(l)})(T_{s}^{(k)}-\mu _ {T_{s}})(T_{i}^{(l)}-\mu _ {T_{i}})/(\sigma _{T_{s}}\sigma _{T_{i}})$ where $\mu _ {T_{s}},\mu _{T_{i}}$ are the mean arrival times of the signal and idler photons and $\sigma _{T_{s}},\sigma _{T_{i}}$ are their standard deviations. We obtain $PCC=0.51\pm0.012$, which indicates a strong correlation. The main source of correlation degradation in our system is the $400 ps$ jitter of the detectors, which causes a circular smearing of the histogram peaks. Another source of decorrelation is the uncertainty in the creation times of the pairs, which results in a diagonal spread of about $\approx 200 ps$ that hardly effects the PCC between the chosen arrival times. In principle, inter modal coupling can also add decorrelation, however the PCC is sensitive only to mode mixing that occurs in the first few tens of centimeters of the fiber, because the arrival times of photons which experience mode coupling after a longer distance will be different from the post selected times $(T_{s}^{(1)},T_{s}^{(2)}),(T_{i}^{(1)},T_{i}^{(2)})$. Thus PCC degradation due to inter modal mode mixing, which is typically on the order of $20db/km$ \cite{kaliteevskiy2013two}, is negligible. 

\subsection{Modal Group Delay Simulation}
\begin{figure}[H]
\begin{centering}
\includegraphics[width=\columnwidth]{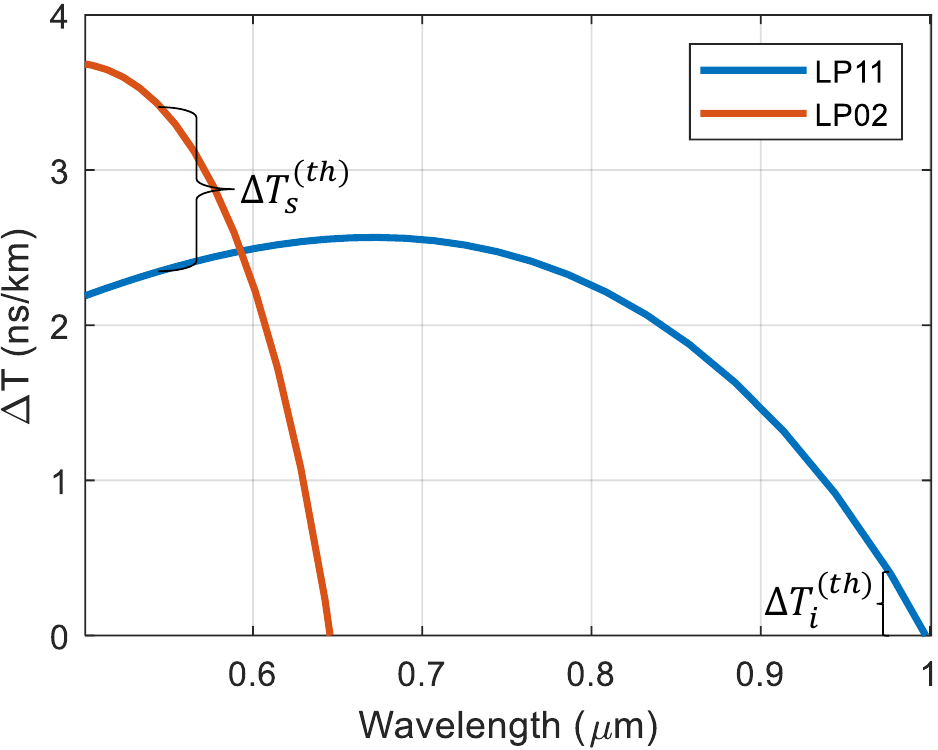}
\par\end{centering}
    \caption{ \textbf{Numerical computation of the modal delays of the $LP_{11}$ (blue curve) and the $LP_{02}$ (red curve) modes.} The delays are presented relative to the fundamental mode $LP_{01}$, to compensate for chromatic delay. For the signal photon at $\lambda_s=542nm$, the delay between the $LP_{02}$ and $LP_{11}$  modes is $\Delta T_{s}^{(th)}=1ns/km$, in agreement with the experimentally measured delays presented in Figure \ref{Fig2}(c). For the idler photon at $ \lambda_i=970nm$, the delay between the $ LP_{11}$ and $LP_{01}$ modes is $\Delta T_{i}^{(th)}=0.5ns/km$, in agreement with measured delays reported in Figure \ref{Fig2}(b).}
\label{Fig3}
\end{figure}

To show that the measured delays between the signal and idler photons match the expected delays for an SMF-28 fiber, we numerically calculated its modal group delays. We solve the scalar wave equation for an SMF-28 fiber, with a $4.2um$ core radius, core-cladding index difference of $\Delta=0.33\%$, and a step-index profile with a typical dip shape. The modal delay of $LP_{11},LP_{02}$ modes, relative to the fundamental mode is presented in Figure \ref{Fig3}. We chose the fundamental mode as a reference to cancel the chromatic dispersion. At the signal's wavelength the delay between the $LP_{02}$ and $LP_{11}$ is $\Delta T_s=1 ns$. At the idler's wavelength the delay of $LP_{01}$ and $LP_{11}$ is $\Delta T_i=0.5 ns$. These delays are in agreement with the temporal correlations found experimentally, supporting mode-to-time mapping scheme. 

\section{Discussion}
 In conclusion, we have demonstrated generation and sorting of correlated photon pairs occupying high order modes of a commercially available fiber. The all-fiber configuration opens the door for implementing high-dimensional photonic quantum bits in fiber-based applications. For example, the mode-to-time mapping can potentially solve the challenge of scaling the number of required detectors with the number of fiber modes, an outstanding challenge in conventional mode sorters. Towards this end, it is necessary to improve the temporal resolution of the system, for example by using superconducting nanowire single photon detectors with jitter times as low as a few picoseconds, and faster electronics. It will allow sorting more transverse modes and using shorter fibers for the temporal mode sorter, which in turn will decrease the background noise caused by fluorescence and parasitic nonlinear processes in the fiber. 
 
In order to manipulate the photons coherently and apply projective measurement in two mutually unbiased bases one can use multi-plane light converters (MPLC) \cite{lib2021reconfigurable,fontaine2019laguerre, fontaine2021hermite,krenn2014generation}. We note that by combining the all-fiber temporal sorter with an all-fiber wavefront modulator that we recently developed \cite{resisi2020wavefront}, it would be possible to demonstrate an all-fiber sorter in mutually unbiased basis, opening the door for all fiber quantum communication protocols with high dimensional quantum bits. 
 
Addressing these challenges will allow exploring applications of the all-fiber source and sorter. For example, using an in-line multimode fiber beam splitter one could split the photon pairs and route each photon to a different remote user. Such configuration is relevant for device independent quantum key distribution, where an untrusted user (Charlie) distributes entangled photon pairs to Alice and Bob, who generate a secure key based on Bell measurements \cite{ekert1991quantum}. A more immediate application of the all-fiber source is quantum communication protocols that rely on sending both photons to the same target. Examples include quantum dense coding \cite{hu2018beating,guo2019advances}, high capacity quantum key distribution \cite{cabello2000quantum,long2002theoretically} and direct quantum communication \cite{deng2003two}. 
\newline

\section{Methods}
\subsection{Experimental Setup}

An optical fiber (SMF-28) is pumped by a Ti:Sapphire laser (Coherent Chameleon Ultra II, 680-1060nm, 140fs duration, 80MHz repetition rate). Before coupling to the fiber, the laser was filtered using a bandpass filter (Thorlabs FB700-40). The signal and idler photons were separated using a dichroic mirror (DM) with an edge at 925 nm (Semrock FF925-Di01). In each arm the pump beam was blocked using spectral filters. In the signal arm we employed a short pass filter (Semrock BSP01-633R), and a bandpass filter (Semrock FF01-540). In the Idler arm we employed a long pass filter (Semrock BLP01-808R) and a bandpass filter (Semrock LL01-976). The signal and idler photons were coupled into two optical fibers (SMF-28) and detected using avalanche photodetectors (Excelitas SPCM-AQ4C), with a quantum efficiency of 50\% for the signal photons and 15\% for the idler photons. The arrival times of the photons where registered using a time-to-digital converter (Swabian Time Tagger 20).

\section{Acknowledgments}
The authors kindly thank Hagai Eisenberg and Avi Pe'er for many fruitful discussions and suggestions. 
This research was supported by the \textit{United States-Israel Binational Science Foundation (BSF)} (Grant No. 2017694). KS and YB acknowledge the support of the Israeli Council for Higher Education, the Israel National Quantum Initiative (INQI) and the Zuckerman STEM Leadership Program. 

\end{multicols}
\clearpage

\maketitle

\section{Supplementary Information}
\subsection{Note 1: Scheme for higher-dimensional state generation at telecom and visible wavelengths}

To generate high-dimensional states in a superposition of multiple fiber modes, it is required to find the phase matching condition that will allow multiple spontaneous four-wave mixing processes in the same spectral channel. Here we show a concrete example of such phase-matching condition based on a commercially available graded index (GRIN) fiber (OM4) spliced to a commercial Ytterbium mode-locked fiber laser. In addition to the possibility to generate high-dimensional quantum states, this scheme also allows generation of photons in the c-band.  

We start by presenting the phase matching condition that allows high-dimensional entanglement in OM4 fibers. In multimode fibers, one can find multiple modal configurations that satisfy the phase matching condition required for four-wave mixing. There are a few benefits in using GRIN fibers. The guided-modes in a GRIN fiber can propagate with nearly identical group velocities and therefore nonlinear coupling among short pulses is achieved over much longer distances than in step index fibers. More importantly for generating high-dimensional quantum bits, the parabolic refractive index profile of GRIN fibers yields degenerate group modes with equally spaced propagation constants $\beta_n$ and a degeneracy that scales linearly with the group number $g_n$. It is therefore possible to obtain multiple combinations of guided-modes that satisfy the phase matching condition for the same signal and idler frequencies  $\omega_s, \omega_i$. Explicitly, the dependence of the phase matched signal and idler frequencies on the group number mismatch defined by $G=-2g_{pump}+g_{idler}+g_{signal}$ is given by \cite{nazemosadat2016phase}:

\begin{equation}
    \omega_{s,i} =\omega_p\pm\sqrt{\frac{\sqrt{2\Delta}G}{R\beta_p^{(2)}(\omega_p)}} 
\end{equation}
Where the $+ (-)$ sign corresponds to the signal (idler) frequency, $\omega_p$ is the pump frequency, $\beta_p^{(2)}(\omega_p)$ is the group-velocity dispersion parameter of the pump mode, and $R$, $\Delta$ are the core radius and the maximal refractive index difference between the core and the clad, respectively. The dimension of the signal and idler photons therefore increases with the group number mismatch $G$. For example, for $G=2$ there are four different modal combinations that yield the same phase matched frequencies, resulting in four-dimensional quantum states at $\omega_s$ and $\omega_i$. 

To confirm the above phase matching analysis, we numerically solve the multimode nonlinear Schrödinger equation (MM-NLSE) using the numerical solver developed in \cite{wright2017multimode}. The MM-NLSE is given by:
\begin{equation}
     \frac{\partial A_{k}}{\partial z} = i\sum_{n} \frac{\beta_n^{(k)}}{n!}(i\frac{\partial}{\partial t})^n A_k + i\frac{n_2\omega_p}{c}\sum_{lmn}S_{klmn}A_lA_mA^*_n
\end{equation}
where $A_k(z,t)$ is the slowly varying amplitude of mode $k$, $z$ is the propagation axis along the fiber and $\beta_n^{(k)} = \partial^n\beta^{(k)}/\partial\omega^n$. The nonlinear coupling coefficients $S_{klmn}$ are given by the overlap of the transverse profiles of the guided-modes $F_k(x,y)$: 
\begin{equation}
S_{klmn} = \frac{\int dxdy F_k^*(x,y)F_n^*(x,y)F_m(x,y)F_l(x,y)}{\sqrt{\int dxdy |F_k(x,y)|^2 \int dxdy |F_l(x,y)|^2 \int dxdy |F_m(x,y)|^2 \int dxdy |F_n(x,y)|^2}}
\end{equation}

\begin{figure}[ht!]
\begin{centering}
\includegraphics[width=0.7\columnwidth]{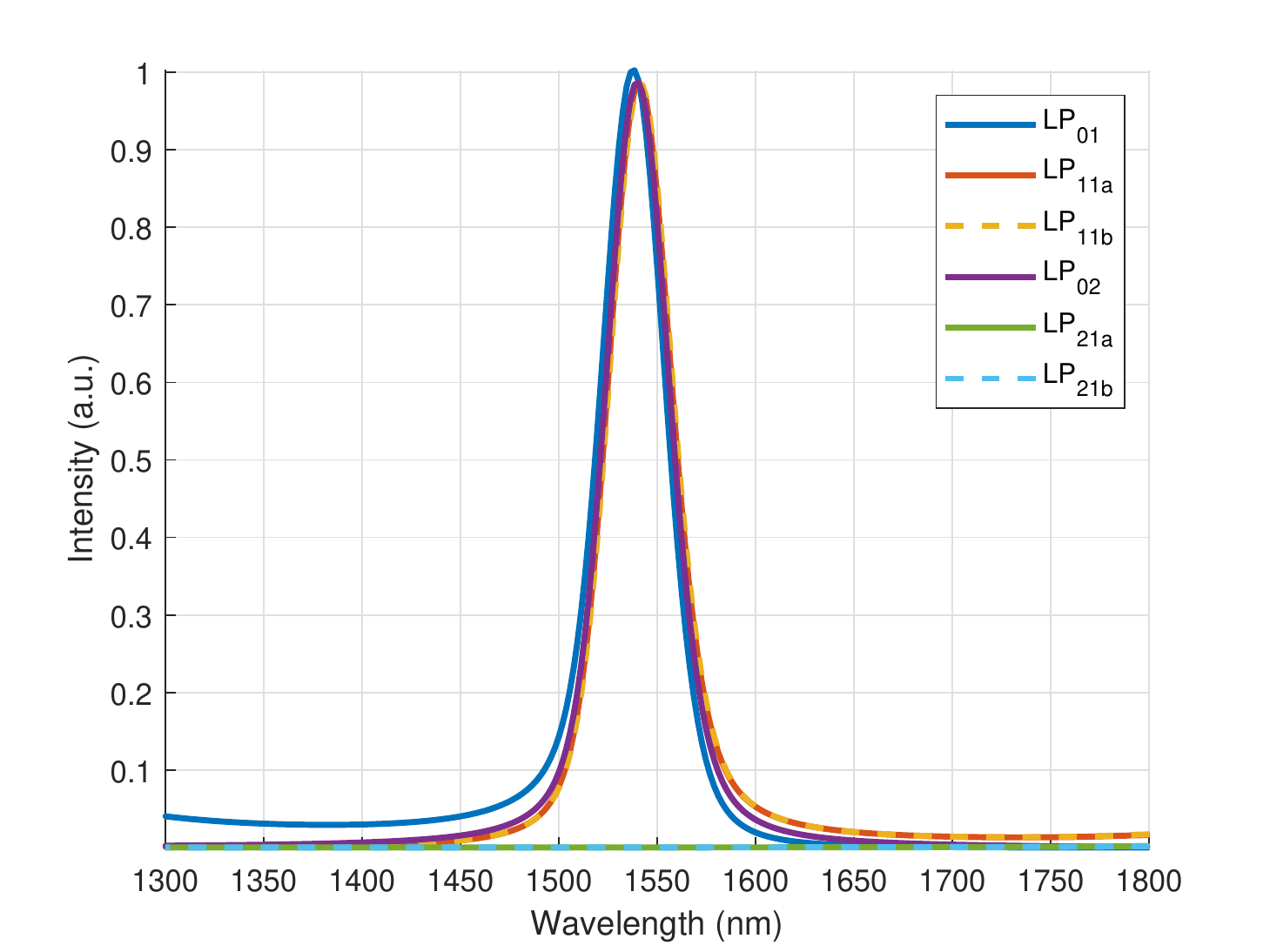}
\par\end{centering}
    \caption{Power spectrum of intermodal four-wave mixing at the output of a 10cm graded index fiber pumped by $140fs$ pulses at $\lambda_p=1040nm$, obtained by numerical integration of the multimode nonlinear Schrödinger equation (MM-NLSE). To find the phase matching wavelengths, all the guided-modes of the fiber at wavelengths lower than $\lambda_p$ were excited.}
\label{SFig1}
\end{figure}

To obtain the phase matched frequencies, we propagate a strong pump field at the fundamental mode of a GRIN  fiber at $\lambda_p=1040nm$, together with a weak signal seed occupying all the guided-modes of the fiber and all wavelengths lower than $\lambda_p$. To simulate a concrete scheme, we use $140fs$ pulses with an energy of $0.1nJ$ per pulse, corresponding to commercially available Ytterbium mode-locked fiber lasers. Supplementary Figure \ref{SFig1} presents the spectrum at the output of a 10cm long fiber, exhibiting idler photons centered at $\lambda=1540nm$ that occupy four fiber modes.

In Supplementary Table \ref{tab1} we summarize the spontaneous four-wave mixing processes that are stimulated by the seed parameters given above, showing that four-dimensional quantum states can be generated in the c-band, by pumping a GRIN fiber with a commercially available femtosecond fibre laser. 

\begin{table}[h!]
\centering
\begin{tabular}{||c c c c c||} 
 \hline
 Signal mode & Idler mode & $\lambda_{Signal}$ (nm) & $\lambda_{Idler}$ (nm) & G \\
 \hline\hline
 $LP_{01}$ & $LP_{02}$ & 785 & 1540 & 2 \\ 
 \hline
 $LP_{02}$ & $LP_{01}$ & 785 & 1540 & 2 \\
 \hline
 $LP_{11a}$ & $LP_{11a}$ & 785 & 1540 & 2 \\
 \hline
 $LP_{11b}$ & $LP_{11b}$ & 785 & 1540 & 2 \\
 \hline
\end{tabular}
\caption{ Phase-matching of intermodal four-wave mixing in a GRIN fiber (OM4). The pump is assumed to be in the $LP_{01}$ mode at $\lambda_p=1040nm$ . At $\lambda_s=785nm$ and $\lambda_i=1540nm$ we get four types of intermodal processes, enabling the generation of a four-dimensional quantum state.}\label{tab1}
\end{table}

The spectral band of the generated photons depends on the pump wavelength. It is possible to generate spontaneous intermodal four wave mixing up to the telecom wavelenghts, thanks to the fact that the phase matching conditions are nearly unaffected by the pump wavelength. For example, in the above table we present a process where the idler is generated in the telecom c-band spectrum. However, because the four-wave mixing process conserves energy, the second photon is generated in 785nm. Such configuration is especially relevant when one wishes to send one photon in free-space and the other photon in a fiber.

\newpage
\subsection{Note 2: Numerical calculation of the pair generation rate}

To analyze the pair generation rate as a function of the fiber length, we numerically integrate the MM-NLSE as described in the previous section,  now for the fibre and pump parameters used in our experiment. We propagate a strong pump pulse occupying the fundamental mode, along with weak white noise occupying all guided-modes that simulates seeding by vacuum fluctuations.  For the pump field we assume $140 fs$, $0.1nJ$ pulses centered at $\lambda_p=695nm$, with a repetition rate of $80MHz$, corresponding to the pulses used in our experiment. For the vacuum fluctuations we assume zero mean fields with nonzero variance which corresponds to an energy of $\hbar\omega$ per spectral channel \cite{trajtenberg2020simulating}. The obtained pair rate at the output of the fiber as a function of the fiber length is presented in Supplementary Figure \ref{SFig2}, showing that most of the photons are generated in the first tens of centimeters of the fiber. We attribute most of the reduction in the pair rate to the pump dispersion, as the dispersion length for the pump pulses, defined by $L_D=T_0^2/\beta_2$ where $T_0=140fs$ is the pulse duration and $\beta_2=44000 fs^2$ is the group velocity dispersion parameter, is approximately 45cm. 

We further note that an upper bound on the fiber segment length over which the photons are generated, can be estimated from the temporal two-photon probability presented in Figure 2 of the main text. Since the temporal signal-idler separation scales linearly with the distance they propagate in the fiber, generation at different positions along the fiber exhibits smearing of the two-photon probability peaks along its diagonal. Figure 2 exhibits a diagonal spreading of $\approx 200ps$. Since the measured signal-idler separation after 1km is $70ns$, the $200ps$ spreading sets an upper of a few meters on the segment length over which the pairs are generated. In practice, since other noise sources may contribute to the diagonal spreading of the two-photon probability, the actual segment length is most likely shorter than this upper bound. 

\begin{figure}[h!]
\begin{centering}
\includegraphics[width=0.7\columnwidth]{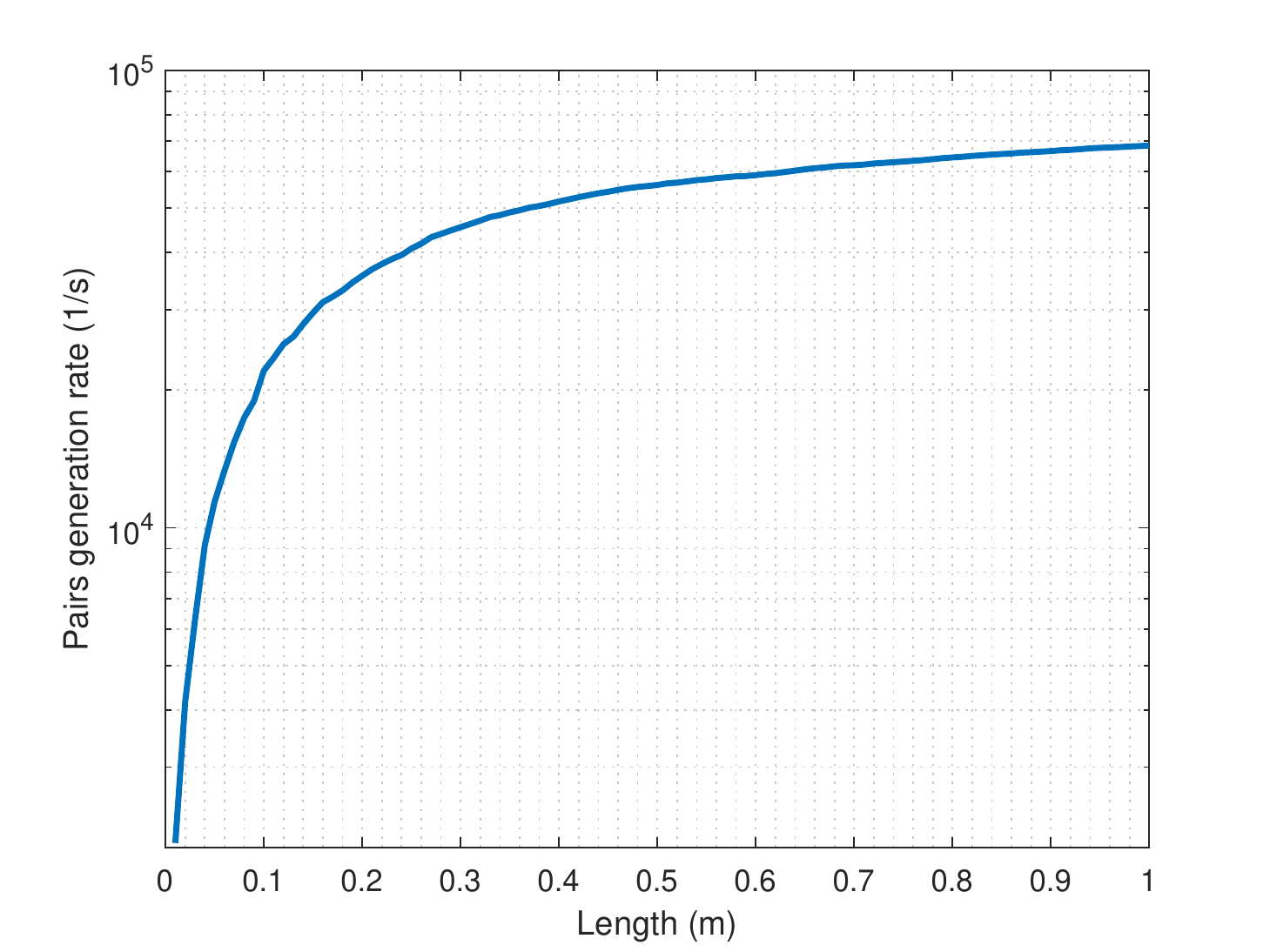}
\par\end{centering}
   \caption{Pair generation rate obtained by numerical integration of the multimode nonlinear Schrödinger equation. Here we propagate $0.1nJ$, 140fs long pump pulses at $\lambda_p=695nm$, together with white noise fields at $\lambda_{vacuum}<695nm$ corresponding to vacuum fluctuations, which have a zero mean and a nonzero variance corresponding to an energy of $\hbar\omega$ per spectral channel. }
\label{SFig2}
\end{figure}

\subsection{Note 3: Coincidence to accidental ratio measurement}

\begin{figure}[h!]
\centering
\includegraphics[width=0.7\columnwidth]{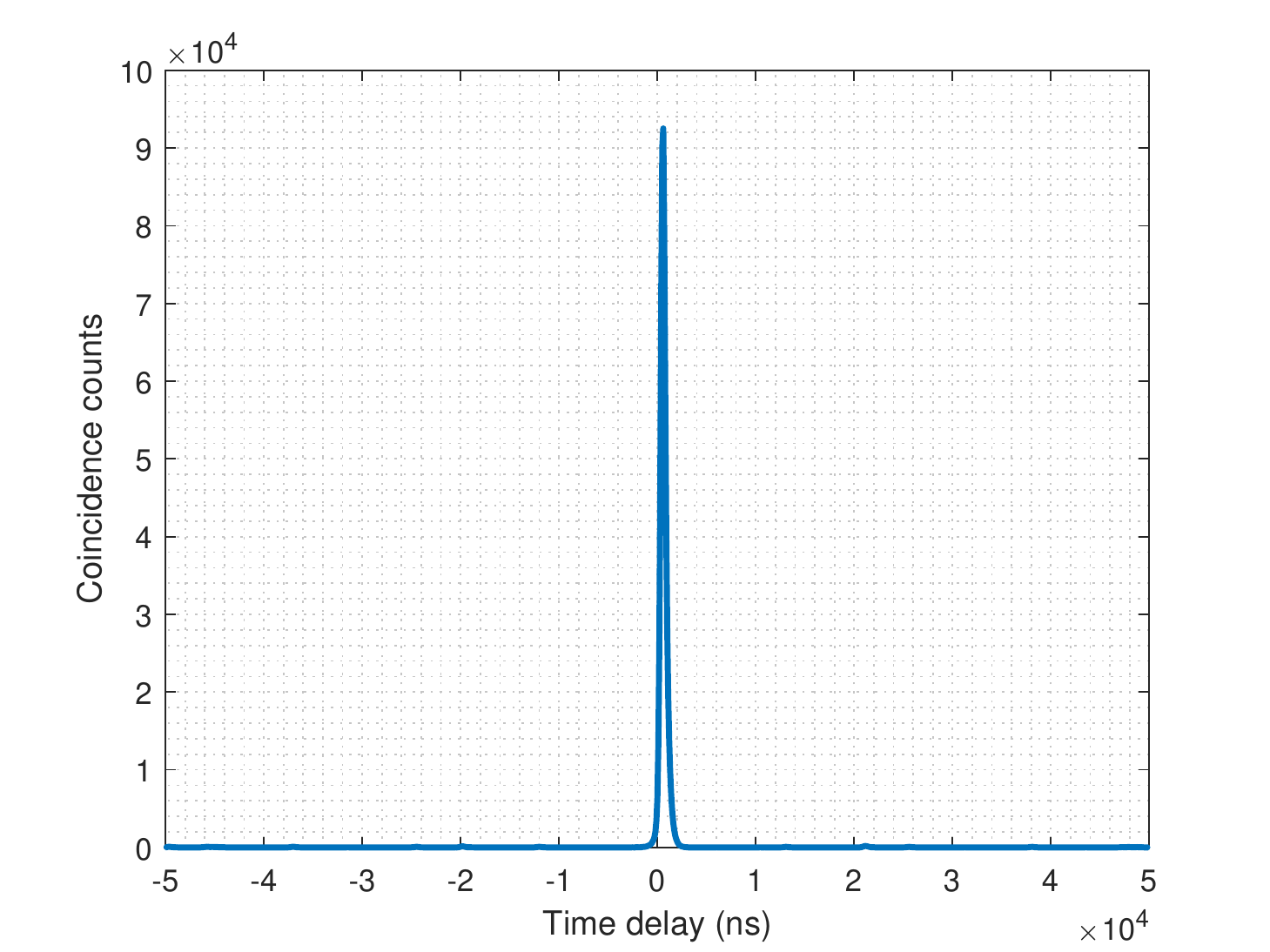}
    \caption{Coincidence histogram ranging a few times the separation between consecutive pump pulses, yielding a coincidence to accidental ratio of $CAR=850$ for a pump average power of $P=10mW$. }
\label{SFig3}
\end{figure}

To quantify the coincidence to accidental ratio (CAR) we measured the coincidence histogram below for a pump average power of $P=10mW$, using 20cm long section of a SMF-28 (Supplementary Figure \ref{SFig3}). The CAR is found by the ratio of the correlation peak to the highest correlation measured at a delay of an integer number of pump periods, $CAR=850$. 

\clearpage
\printbibliography

@article{cerf2002security,
  title={Security of quantum key distribution using d-level systems},
  author={Cerf, Nicolas J and Bourennane, Mohamed and Karlsson, Anders and Gisin, Nicolas},
  journal={Physical review letters},
  volume={88},
  number={12},
  pages={127902},
  year={2002},
  publisher={APS}
}

@article{erhard2020advances,
  title={Advances in high-dimensional quantum entanglement},
  author={Erhard, Manuel and Krenn, Mario and Zeilinger, Anton},
  journal={Nature Reviews Physics},
  pages={1--17},
  year={2020},
  publisher={Nature Publishing Group}
}

@article{erhard2018twisted,
  title={Twisted photons: new quantum perspectives in high dimensions},
  author={Erhard, Manuel and Fickler, Robert and Krenn, Mario and Zeilinger, Anton},
  journal={Light: Science \& Applications},
  volume={7},
  number={3},
  pages={17146--17146},
  year={2018},
  publisher={Nature Publishing Group}
}

@article{molina2007twisted,
  title={Twisted photons},
  author={Molina-Terriza, Gabriel and Torres, Juan P and Torner, Lluis},
  journal={Nature physics},
  volume={3},
  number={5},
  pages={305--310},
  year={2007},
  publisher={Nature Publishing Group}
}

@article{krenn2015twisted,
  title={Twisted photon entanglement through turbulent air across Vienna},
  author={Krenn, Mario and Handsteiner, Johannes and Fink, Matthias and Fickler, Robert and Zeilinger, Anton},
  journal={Proceedings of the National Academy of Sciences},
  volume={112},
  number={46},
  pages={14197--14201},
  year={2015},
  publisher={National Acad Sciences}
}

@article{sit2017high,
  title={High-dimensional intracity quantum cryptography with structured photons},
  author={Sit, Alicia and Bouchard, Fr{\'e}d{\'e}ric and Fickler, Robert and Gagnon-Bischoff, J{\'e}r{\'e}mie and Larocque, Hugo and Heshami, Khabat and Elser, Dominique and Peuntinger, Christian and G{\"u}nthner, Kevin and Heim, Bettina and others},
  journal={Optica},
  volume={4},
  number={9},
  pages={1006--1010},
  year={2017},
  publisher={Optical Society of America}
}

@article{loffler2011fiber,
  title={Fiber transport of spatially entangled photons},
  author={L{\"o}ffler, W and Euser, TG and Eliel, ER and Scharrer, M and Russell, P St J and Woerdman, JP},
  journal={Physical review letters},
  volume={106},
  number={24},
  pages={240505},
  year={2011},
  publisher={APS}
}

@article{kang2012measurement,
  title={Measurement of the entanglement between photonic spatial modes in optical fibers},
  author={Kang, Yoonshik and Ko, Jaekwon and Lee, Sang Min and Choi, Sang-Kyung and Kim, Byoung Yoon and Park, Hee Su},
  journal={Physical review letters},
  volume={109},
  number={2},
  pages={020502},
  year={2012},
  publisher={APS}
}

@article{cozzolino2019air,
  title={Air-core fiber distribution of hybrid vector vortex-polarization entangled states},
  author={Cozzolino, Daniele and Polino, Emanuele and Valeri, Mauro and Carvacho, Gonzalo and Bacco, Davide and Spagnolo, Nicol{\`o} and Oxenl{\o}we, Leif K and Sciarrino, Fabio},
  journal={Advanced Photonics},
  volume={1},
  number={4},
  pages={046005},
  year={2019},
  publisher={International Society for Optics and Photonics}
}

@article{valencia2020unscrambling,
  title={Unscrambling entanglement through a complex medium},
  author={Valencia, Natalia Herrera and Goel, Suraj and McCutcheon, Will and Defienne, Hugo and Malik, Mehul},
  journal={Nature Physics},
  pages={1--5},
  year={2020},
  publisher={Nature Publishing Group}
}

@article{cao2020distribution,
  title={Distribution of high-dimensional orbital angular momentum entanglement over a 1 km few-mode fiber},
  author={Cao, Huan and Gao, She-Cheng and Zhang, Chao and Wang, Jian and He, De-Yong and Liu, Bi-Heng and Zhou, Zheng-Wei and Chen, Yu-Jie and Li, Zhao-Hui and Yu, Si-Yuan and others},
  journal={Optica},
  volume={7},
  number={3},
  pages={232--237},
  year={2020},
  publisher={Optical Society of America}
}

@article{cruz2016fiber,
  title={Fiber-based photon-pair source capable of hybrid entanglement in frequency and transverse mode, controllably scalable to higher dimensions},
  author={Cruz-Delgado, D and Ramirez-Alarcon, R and Ortiz-Ricardo, E and Monroy-Ruz, J and Dominguez-Serna, F and Cruz-Ramirez, H and Garay-Palmett, K and U’Ren, AB},
  journal={Scientific reports},
  volume={6},
  pages={27377},
  year={2016},
  publisher={Nature Publishing Group}
}

@article{rottwitt2018photon,
  title={Photon-pair sources based on intermodal four-wave mixing in few-mode fibers},
  author={Rottwitt, Karsten and Koefoed, Jacob Gade and Christensen, Erik Nicolai},
  journal={Fibers},
  volume={6},
  number={2},
  pages={32},
  year={2018},
  publisher={Multidisciplinary Digital Publishing Institute}
}

@article{guo2019generation,
  title={Generation of telecom-band correlated photon pairs in different spatial modes using few-mode fibers},
  author={Guo, Cheng and Su, Jie and Zhang, Zhenzhen and Cui, Liang and Li, Xiaoying},
  journal={Optics Letters},
  volume={44},
  number={2},
  pages={235--238},
  year={2019},
  publisher={Optical Society of America}
}

@inproceedings{shamsshooli2020toward,
  title={Toward Generation of Spatially-Entangled Photon Pairs in a Few-Mode Fiber},
  author={Shamsshooli, Afshin and Guo, Cheng and Parmigiani, Francesca and Li, Xiaoying and Vasilyev, Michael},
  booktitle={CLEO: Applications and Technology},
  pages={JTh2A--27},
  year={2020},
  organization={Optical Society of America}
}

@article{pourbeyram2018photon,
  title={Photon pair generation with tailored frequency correlations in graded-index multimode fibers},
  author={Pourbeyram, Hamed and Mafi, Arash},
  journal={Optics letters},
  volume={43},
  number={9},
  year={2018},
  publisher={Optical Society of America}
}

@article{rottwitt2019quantum,
  title={Quantum information processing using intermodal four-wave mixing in multi-mode optical fibers},
  author={Rottwitt, Karsten and Christensen, Jesper B and Christensen, Erik N and Koefoed, Jacob G},
  journal={2019 21st International Conference on Transparent Optical Networks (ICTON)},
  pages={1--3},
  year={2019},
  organization={IEEE}
}

@article{ekici2020graded,
  title={Graded-index optical fiber transverse-spatial-mode entanglement},
  author={Ekici, Cagin and Dinleyici, Mehmet Salih},
  journal={Physical Review A},
  volume={102},
  number={1},
  pages={013702},
  year={2020},
  publisher={APS}
}

@article{goudreau2020theory,
  title={Theory of four-wave mixing of cylindrical vector beams in optical fibers},
  author={Goudreau, E Scott and Kupchak, Connor and Sussman, Benjamin J and Boyd, Robert W and Lundeen, Jeff S},
  journal={JOSA B},
  volume={37},
  number={6},
  pages={1670--1682},
  year={2020},
  publisher={Optical Society of America}
}

@article{fontaine2019laguerre,
  title={Laguerre-Gaussian mode sorter},
  author={Fontaine, Nicolas K and Ryf, Roland and Chen, Haoshuo and Neilson, David T and Kim, Kwangwoong and Carpenter, Joel},
  journal={Nature communications},
  volume={10},
  number={1},
  pages={1--7},
  year={2019},
  publisher={Nature Publishing Group}
}

@article{berkhout2010efficient,
  title={Efficient sorting of orbital angular momentum states of light},
  author={Berkhout, Gregorius CG and Lavery, Martin PJ and Courtial, Johannes and Beijersbergen, Marco W and Padgett, Miles J},
  journal={Physical review letters},
  volume={105},
  number={15},
  pages={153601},
  year={2010},
  publisher={APS}
}

@article{mair2001entanglement,
  title={Entanglement of the orbital angular momentum states of photons},
  author={Mair, Alois and Vaziri, Alipasha and Weihs, Gregor and Zeilinger, Anton},
  journal={Nature},
  volume={412},
  number={6844},
  pages={313--316},
  year={2001},
  publisher={Nature Publishing Group}
}

@article{krenn2014generation,
  title={Generation and confirmation of a (100$\times$ 100)-dimensional entangled quantum system},
  author={Krenn, Mario and Huber, Marcus and Fickler, Robert and Lapkiewicz, Radek and Ramelow, Sven and Zeilinger, Anton},
  journal={Proceedings of the National Academy of Sciences},
  volume={111},
  number={17},
  pages={6243--6247},
  year={2014},
  publisher={National Acad Sciences}
}

@article{li2005optical,
  title={Optical-fiber source of polarization-entangled photons in the 1550 nm telecom band},
  author={Li, Xiaoying and Voss, Paul L and Sharping, Jay E and Kumar, Prem},
  journal={Physical review letters},
  volume={94},
  number={5},
  pages={053601},
  year={2005},
  publisher={APS}
}

@article{takesue20051,
  title={1.5-$\mu$m band quantum-correlated photon pair generation in dispersion-shifted fiber: suppression of noise photons by cooling fiber},
  author={Takesue, Hiroki and Inoue, Kyo},
  journal={Optics express},
  volume={13},
  number={20},
  pages={7832--7839},
  year={2005},
  publisher={Optical Society of America}
}

@article{dyer2009high,
  title={High-brightness, low-noise, all-fiber photon pair source},
  author={Dyer, Shellee D and Baek, Burm and Nam, Sae Woo},
  journal={Optics express},
  volume={17},
  number={12},
  pages={10290--10297},
  year={2009},
  publisher={Optical Society of America}
}

@article{sharping2004quantum,
  title={Quantum-correlated twin photons from microstructure fiber},
  author={Sharping, Jay E and Chen, Jun and Li, Xiaoying and Kumar, Prem and Windeler, Robert S},
  journal={Optics Express},
  volume={12},
  number={14},
  pages={3086--3094},
  year={2004},
  publisher={Optical Society of America}
}

@article{rarity2005photonic,
  title={Photonic crystal fiber source of correlated photon pairs},
  author={Rarity, JG and Fulconis, J and Duligall, J and Wadsworth, WJ and Russell, P St J},
  journal={Optics express},
  volume={13},
  number={2},
  pages={534--544},
  year={2005},
  publisher={Optical Society of America}
}

@article{fan2005efficient,
  title={Efficient generation of correlated photon pairs in a microstructure fiber},
  author={Fan, Jingyun and Migdall, Alan and Wang, LJ},
  journal={Optics letters},
  volume={30},
  number={24},
  pages={3368--3370},
  year={2005},
  publisher={Optical Society of America}
}

@article{cohen2009tailored,
  title={Tailored photon-pair generation in optical fibers},
  author={Cohen, Offir and Lundeen, Jeff S and Smith, Brian J and Puentes, Graciana and Mosley, Peter J and Walmsley, Ian A},
  journal={Physical review letters},
  volume={102},
  number={12},
  pages={123603},
  year={2009},
  publisher={APS}
}

@article{smith2009photon,
  title={Photon pair generation in birefringent optical fibers},
  author={Smith, Brian J and Mahou, P and Cohen, Offir and Lundeen, JS and Walmsley, IA},
  journal={Optics express},
  volume={17},
  number={26},
  pages={23589--23602},
  year={2009},
  publisher={Optical Society of America}
}

@article{lugani2020spectrally,
  title={Spectrally pure single photons at telecommunications wavelengths using commercial birefringent optical fiber},
  author={Lugani, Jasleen and Francis-Jones, Robert JA and Boutari, Joelle and Walmsley, Ian A},
  journal={Optics Express},
  volume={28},
  number={4},
  pages={5147--5163},
  year={2020},
  publisher={Optical Society of America}
}

@article{chandrasekharan2020observing,
  title={Observing mode-dependent wavelength-to-time mapping in few-mode fibers using a single-photon detector array},
  author={Chandrasekharan, Harikumar K and Ehrlich, Katjana and Tanner, Michael G and Haynes, Dionne M and Mukherjee, Sebabrata and Birks, Tim A and Thomson, Robert R},
  journal={APL Photonics},
  volume={5},
  number={6},
  pages={061303},
  year={2020},
  publisher={AIP Publishing LLC}
}

@article{painchaud1992time,
  title={Time-resolved identification of modes and measurement of intermodal dispersion in optical fibers},
  author={Painchaud, Yves and LeBel, P and Duguay, MA and Black, Richard J},
  journal={Applied optics},
  volume={31},
  number={12},
  pages={2005--2010},
  year={1992},
  publisher={Optical Society of America}
}

@article{park2020telecom,
  title={Telecom C-band photon-pair generation using standard SMF-28 fiber},
  author={Park, Kyungdeuk and Lee, Dongjin and Boyd, Robert W and Shin, Heedeuk},
  journal={Optics Communications},
  pages={126692},
  year={2020},
  publisher={Elsevier}
}

@article{kaliteevskiy2013two,
  title={Two-mode coupling model in a few mode fiber},
  author={Kaliteevskiy, NA and Korolev, AE and Koreshkov, KS and Nazarov, VN and Sterlingov, PM},
  journal={Optics and Spectroscopy},
  volume={114},
  number={6},
  pages={913--916},
  year={2013},
  publisher={Springer}
}

@article{resisi2020wavefront,
  title={Wavefront shaping in multimode fibers by transmission matrix engineering},
  author={Resisi, Shachar and Viernik, Yehonatan and Popoff, Sebastien M and Bromberg, Yaron},
  journal={APL Photonics},
  volume={5},
  number={3},
  pages={036103},
  year={2020},
  publisher={AIP Publishing LLC}
}

@article{zhou2021high,
  title={High-fidelity spatial mode transmission through a 1-km-long multimode fiber via vectorial time reversal},
  author={Zhou, Yiyu and Braverman, Boris and Fyffe, Alexander and Zhang, Runzhou and Zhao, Jiapeng and Willner, Alan E and Shi, Zhimin and Boyd, Robert W},
  journal={Nature Communications},
  volume={12},
  number={1},
  pages={1--7},
  year={2021},
  publisher={Nature Publishing Group}
}

@article{cozzolino2019high,
  title={High-dimensional quantum communication: benefits, progress, and future challenges},
  author={Cozzolino, Daniele and Da Lio, Beatrice and Bacco, Davide and Oxenl{\o}we, Leif Katsuo},
  journal={Advanced Quantum Technologies},
  volume={2},
  number={12},
  pages={1900038},
  year={2019},
  publisher={Wiley Online Library}
}

@article{alarcon2021few,
  title={Few-mode fibre technology fine-tunes losses of quantum communication systems},
  author={Alarc{\'o}n, A and Argillander, J and Lima, G and Xavier, GB},
  journal={arXiv preprint arXiv:2103.05018},
  year={2021}
}

@inproceedings{fontaine2021hermite,
  title={Hermite-Gaussian mode multiplexer supporting 1035 modes},
  author={Fontaine, Nicolas K and Chen, Haoshuo and Mazur, Mikael and Dallachiesa, Lauren and Kim, KW and Ryf, Roland and Neilson, David and Carpenter, Joel},
  booktitle={2021 Optical Fiber Communications Conference and Exhibition (OFC)},
  pages={1--3},
  year={2021},
  organization={IEEE}
}

@article{lib2021reconfigurable,
  title={Reconfigurable synthesizer for quantum information processing of high-dimensional entangled photons},
  author={Lib, Ohad and Sulimany, Kfir and Bromberg, Yaron},
  journal={arXiv preprint arXiv:2108.02258},
  year={2021}
}

@article{wright2017multimode,
  title={Multimode nonlinear fiber optics: massively parallel numerical solver, tutorial, and outlook},
  author={Wright, Logan G and Ziegler, Zachary M and Lushnikov, Pavel M and Zhu, Zimu and Eftekhar, M Amin and Christodoulides, Demetrios N and Wise, Frank W},
  journal={IEEE Journal of Selected Topics in Quantum Electronics},
  volume={24},
  number={3},
  pages={1--16},
  year={2017},
  publisher={IEEE}
}

@article{nazemosadat2016phase,
  title={Phase matching for spontaneous frequency conversion via four-wave mixing in graded-index multimode optical fibers},
  author={Nazemosadat, Elham and Pourbeyram, Hamed and Mafi, Arash},
  journal={JOSA B},
  volume={33},
  number={2},
  pages={144--150},
  year={2016},
  publisher={Optical Society of America}
}

@article{hu2018beating,
  title={Beating the channel capacity limit for superdense coding with entangled ququarts},
  author={Hu, Xiao-Min and Guo, Yu and Liu, Bi-Heng and Huang, Yun-Feng and Li, Chuan-Feng and Guo, Guang-Can},
  journal={Science advances},
  volume={4},
  number={7},
  pages={eaat9304},
  year={2018},
  publisher={American Association for the Advancement of Science}
}

@article{guo2019advances,
  title={Advances in quantum dense coding},
  author={Guo, Yu and Liu, Bi-Heng and Li, Chuan-Feng and Guo, Guang-Can},
  journal={Advanced Quantum Technologies},
  volume={2},
  number={5-6},
  pages={1900011},
  year={2019},
  publisher={Wiley Online Library}
}

@article{cabello2000quantum,
  title={Quantum key distribution in the Holevo limit},
  author={Cabello, Ad{\'a}n},
  journal={Physical Review Letters},
  volume={85},
  number={26},
  pages={5635},
  year={2000},
  publisher={APS}
}

@article{long2002theoretically,
  title={Theoretically efficient high-capacity quantum-key-distribution scheme},
  author={Long, Gui-Lu and Liu, Xiao-Shu},
  journal={Physical Review A},
  volume={65},
  number={3},
  pages={032302},
  year={2002},
  publisher={APS}
}

@article{deng2003two,
  title={Two-step quantum direct communication protocol using the Einstein-Podolsky-Rosen pair block},
  author={Deng, Fu-Guo and Long, Gui Lu and Liu, Xiao-Shu},
  journal={Physical Review A},
  volume={68},
  number={4},
  pages={042317},
  year={2003},
  publisher={APS}
}

@article{ekert1991quantum,
  title={Quantum cryptography based on Bell’s theorem},
  author={Ekert, Artur K},
  journal={Physical review letters},
  volume={67},
  number={6},
  pages={661},
  year={1991},
  publisher={APS}
}

@article{trajtenberg2020simulating,
  title={Simulating Correlations of Structured Spontaneously Down-Converted Photon Pairs},
  author={Trajtenberg-Mills, Sivan and Karnieli, Aviv and Voloch-Bloch, Noa and Megidish, Eli and Eisenberg, Hagai S and Arie, Ady},
  journal={Laser \& Photonics Reviews},
  volume={14},
  number={3},
  pages={1900321},
  year={2020},
  publisher={Wiley Online Library}
}

\end{document}